\documentclass{article}

\usepackage{arxiv}

\usepackage[utf8]{inputenc} 
\usepackage[T1]{fontenc}    
\usepackage{hyperref}       
\usepackage{url}            
\usepackage{booktabs}       
\usepackage{amsfonts}       
\usepackage{nicefrac}       
\usepackage{microtype}      
\usepackage{lipsum}
\usepackage{graphicx}

\title{MetaSecure: A Passwordless Authentication for the Metaverse}

\author{
 Sibi Chakkaravarthy Sethuraman\\
  Centre of Excellence, Artificial Intelligence \& Robotics (AIR),\\
  School of Computer Science and Engineering\\
  VIT-AP University, India \\
  \texttt{sb.sibi@gmail.com} \\
   \And
 Aditya Mitra \\
 Centre of Excellence, Artificial Intelligence \& Robotics (AIR),\\
  School of Computer Science and Engineering\\
  VIT-AP University, India \\
  \texttt{adityaarghya0@gmail.com} \\
 \And
 Anisha Ghosh \\
Centre of Excellence, Artificial Intelligence \& Robotics (AIR),\\
  School of Computer Science and Engineering\\
  VIT-AP University, India \\
 \And
  Gautam Galada \\
  Centre of Excellence, Artificial Intelligence \& Robotics (AIR),\\
  School of Computer Science and Engineering\\
  VIT-AP University, India \\
   \And
  Anitha Subramanian \\
  Centre of Excellence, Artificial Intelligence \& Robotics (AIR),\\
  School of Computer Science and Engineering\\
  VIT-AP University, India \\
}

\begin{document}
\maketitle
\begin{abstract}
Metaverse in general holds a potential future for cyberspace. At the beginning of Web 2.0, it was witnessed that people were signing in with various pseudonyms or ‘nyms’, risking their online identities  by increasing presence of fake accounts leading to difficulty in unique identification for different roles. However, in Web 3.0, the metaverse, a user's identity is tied to their original identity, where risking one poses a significant risk to the other. Therefore, this paper proposes a novel authentication system for securing digital assets, online identity, avatars, and accounts called Metasecure where a unique id for every entity or user to develop a human establishment is essential on a digital platform. The proposed passwordless system provides three layers of security using device attestation, facial recognition and use of physical security keys, security keys, or smartcards in accordance to  Fast IDentity Online (FIDO2) specifications. It provides SDKs for authentication on any system including VR/XR glasses, thus ensuring seamlessness in accessing services in the Metaverse.
\end{abstract}


\section{Introduction}
Metaverse presents a set of virtual reality platforms known as ‘worlds’ for communications, interactions, shared working, gaming, entertainment, and so on. The idea of the Metaverse and its implementation promotes sustainable growth of human civilization which breaks the barriers of remote communication and makes it more reliable for everyone where every entity will be authenticated on a virtual platform, unlike communications via phone calls or meets \cite{ref1}. 

This also brings in the concept of digital assets. Digital assets like NFTs and Cryptocurrencies are inherently tied to the user’s online identity. Much like physical assets, they also cost and need to be protected against theft. Identity theft in the metaverse may also mean the loss of digital assets. Further, since it is also tied to the user’s real identity, it would have consequences in the real world as well. This leaves the attackers or spam callers as outliers from the clusters of authorized people. A wide range of security breaches and privacy invasions may arise in the metaverse from the management of massive data streams, pervasive user profiling activities, unfair outcomes of AI algorithms, to the safety of physical infrastructures and human bodies \cite{ref2}.

The metaverse will be fertile ground for social engineering attacks. Given that users look like avatars in the metaverse, there is an obvious concern that these avatars will be stolen, falsified or manipulated by bad actors \cite{ref3}. After an attacker takes over a user’s avatar, he or she could possibly then request information from that user’s colleagues.  Confirming and log-in the same is important to fight cybercrimes and related issues. Sexual harassment, eve teasing, stalking and similar cybercrimes on a virtual scale become easier for the criminals when their digital  identity is not monitored. According to reports from a 43- year-old British woman, Nina Jane Patel, who is the vice president of Metaverse Research for Kabuni Ventures, she was “verbally and sexually harassed” by three or four male avatars of Meta Horizon Venues \cite{ref4}. A 21-year-old woman had similar reports. There are various more such cases of cyberstalking, harassment and so on by many users. Strongly identifying each user on the metaverse may be able to solve such cybercrimes that may leave a user in trauma and other mental conditions \cite{ref5}.

Hence, it is important to bind a user’s virtual identity to their actual identity to prevent similar cases and aid law enforcement. It is important to make sure no user can create fake or throwaway accounts and get away with such deeds.

Currently, passwords are the most widely used authentication technology for accessing various online services and resources. However, the use of passwords and other memorized secrets has a lot of vulnerabilities, including social engineering, keylogging, phishing, vishing, and so on, making it non-suitable for protecting important digital assets \cite{ref6}. Passwords being captured, shared, or distributed gives a person control of the entire virtual identity of someone else. Virtual reality headsets are more reliant to use to trave rather than the use of traditional keyboards and mice. These headsets prefer the use of passwordless authentications and biometrics as entering passwords are difficult[6]. 

In today’s world, if a colleague’s email is compromised, a phishing attack or nefarious request may come from that compromised email. However, in the metaverse, a request coming from a compromised avatar will likely be harder to spot. It remains to be seen how well colleagues will be able to ascertain the legitimacy of each other’s avatars, and it seems likely that it will be difficult to identify a stolen avatar in the workplace which is why using Fast Identity Online (FIDO2) standards for authenticating as per this paper will ensure only verified user gains entry , identify culprits and decrease such crimes at a manifold \cite{ref7}.

Authenticating a user would just indicate the user has the knowledge of the secrets and/or in possession of the authentication devices. It does not verify the identity of the user. For example, it is indeed possible for malicious users to extract the same information from legitimate users with an array of attacks like social engineering. Some other advanced forms of attacks include session hijacking, Man in the middle (MITM) and so on \cite{ref8}. Hence, it is important to use multi factor authentication methods that are less vulnerable to common attacks. It is generally needed to verify the identity of the user. Hence, it is not very convenient to use passwords in Metaverse as per the current security measures. In this paper we propose device attestation , physical key verification and facial recognition to ensure the user’s identity  which is much more suited from a metaverse point of view.

\section{Background and Related Study}
Metaverse has experienced many cyber-aggressions that initially start on a specific platform or in relation to a specific topic (e.g., a game) can also subsequently expand to other platforms or topics, thus involving additional users and communities, as it happened in the case of the Gamergate campaign \cite{ref10}, or even in Second Life \cite{ref11}, \cite{ref12}. In a metaverse characterized by a multitude of interconnections between communities, spaces, and applications, these risks are inevitably amplified. Similar studies suggest use of passwords and OTPs to gain access to multiverse. However, by a comparative study of the attack surface on various websites and online services, it can be concluded that passwords, OTPs, some cryptographic authenticator devices like Time-based OTP (TOTP) key fobs are severely vulnerable to social engineering and related attacks. 
There are various exiting research models suggesting facial recognition models for secure authentication. However, two mostly exploited vulnerabilities of the same are the use of deepfakes \cite{ref13} and Presentation attacks, defined as “Presentation to the biometric data capture subsystem with the goal of interfering with the operation of the biometric system” \cite{ref14}. Hence even passwordless models result in being vulnerable to attacks. The availability of public datasets containing facial images has helped researchers in developing
architectures that can be used for facial morphing where faces of two different identities can be blended into one, such that it contains attributes of both faces, this will help the attacker to pass the verification using the new facial image proving it to be an insecure method \cite{ref15}.

Due to previous existing vulnerable methods many organizations are implementing FIDO2 in their systems. Existing studies in the literature shows that FIDO UAF and World Wide Web Consortium (W3C) Verifiable Credentials can be used to present a user-centric and decentralized digital identity system \cite{ref16},\cite{ref17}. It has made digital identity highly trustworthy both for the user and the service provider who may be authenticating the user. The entire system was implemented for a banking scenario to show how secure it could be and has also allowed users to generate on-demand identities that could contain only the necessary information \cite{ref18}. Their model presented the service provider with the authenticated information from the source directly. Another paper presented implementation of FIDO2 authentication to secure physical assets. 

Another paper presented the application of FIDO protocol to enable multi-factor authentication in banking scenarios. It allowed a single gesture phishing-resistant multi-factor authentication. It involves the keys and biometrics to stay on the user’s device and no server-side secrets. It also ensures no third party protocol is involved \cite{ref19}. A study proposed a promising approach to maintain security even after a FIDO authentication is done. A continuous FIDO authentication browser extension allows the Relying Party (RP) and the authenticator to continuously exchange verification in the background. It has been validated using an Android-based roaming authenticator communicating via BLE \cite{ref20}. Another research \cite{ref21} presented a large-scale lab study of FIDO2 single-factor authentication and collected insights about the perception, acceptance, and concerns about passwordless authentication among the end-users. Their results showed that users are willing to accept a replacement of text-based passwords with a security key for single-factor authentication \cite{ref22}. Hence FIDO can be claimed as the new standard for passwordless authentication which ensures user traversing the Metaverse is well identifies and genuine.

It has also been tested that Metasecure implementing use of FIDO2 standards to authenticate user is much faster and more seamless to use as shown in table 1 where the total time of processing can be approximately 496.4ms.
 
Table \ref{tab1:FIDO Authentic} does a comparative study between existing authentication methods along with the proposed standard where we are implementing triple layer authentication system.
\begin{table}[!htbp]
	\caption{FIDO Authentication Time Taken}
	\centering
	\label{tab1:FIDO Authentic}       
	\begin{center}
	\begin{tabular}{|l|c|c|c|}
		\hline
		\textbf{Sl.No} & \textbf{Time taken to} & \textbf{Time taken to verify} & \textbf{Total time for}  \\
		& \textbf{create challenge}& \textbf{challenge and} & \textbf{processing}\\
		& & \textbf{authenticate user} & \textbf{processing}\\
		\hline
		1. & 212 ms & 319 ms & 531 ms \\
		\hline
		2. & 193 ms & 203 ms & 396 ms \\
		\hline
		3. & 224 ms & 324 ms & 548 ms \\
		\hline
		4. & 234 ms & 304 ms & 538 ms\\
		\hline
		5. & 260 ms & 209 ms & 469 ms\\
		\hline
		Average & 224.6 ms & 271.8 ms & 496.4 ms\\
		\hline
	\end{tabular}
\end{center}
\end{table}

\begin{table*}[!htbp]
	\caption{Comparing Metasecure with other recent Metaverse Authentication methods }
	\centering
	\label{tab2:Comparing Metasecure}       
	\begin{center}
	\begin{tabular}{|l|c|c|}
		\hline
		\textbf{Authentication model} & \textbf{Time taken} & \textbf{Security}\\
		\hline
		Password Authentication & 1056.4 ms & Low  \\
		\hline
		Facial recognition Authentication & 154.5 ms & Medium  \\
		\hline
		Passwordless Authentication & 496.4 ms & High  \\
		\hline
		\textbf{Metasecure(FIDO Authentication + device attestation + facial recognition)} &\textbf{128 + 496.4 = 624.4 ms} &\textbf{Extremely high } \\
		
		\hline
	\end{tabular}
\end{center}
\end{table*}

\section{Proposed Solution and Novelty of Metasecure}

\begin{figure*}[!htbp]
	\centering
	\includegraphics[width=0.9\linewidth]{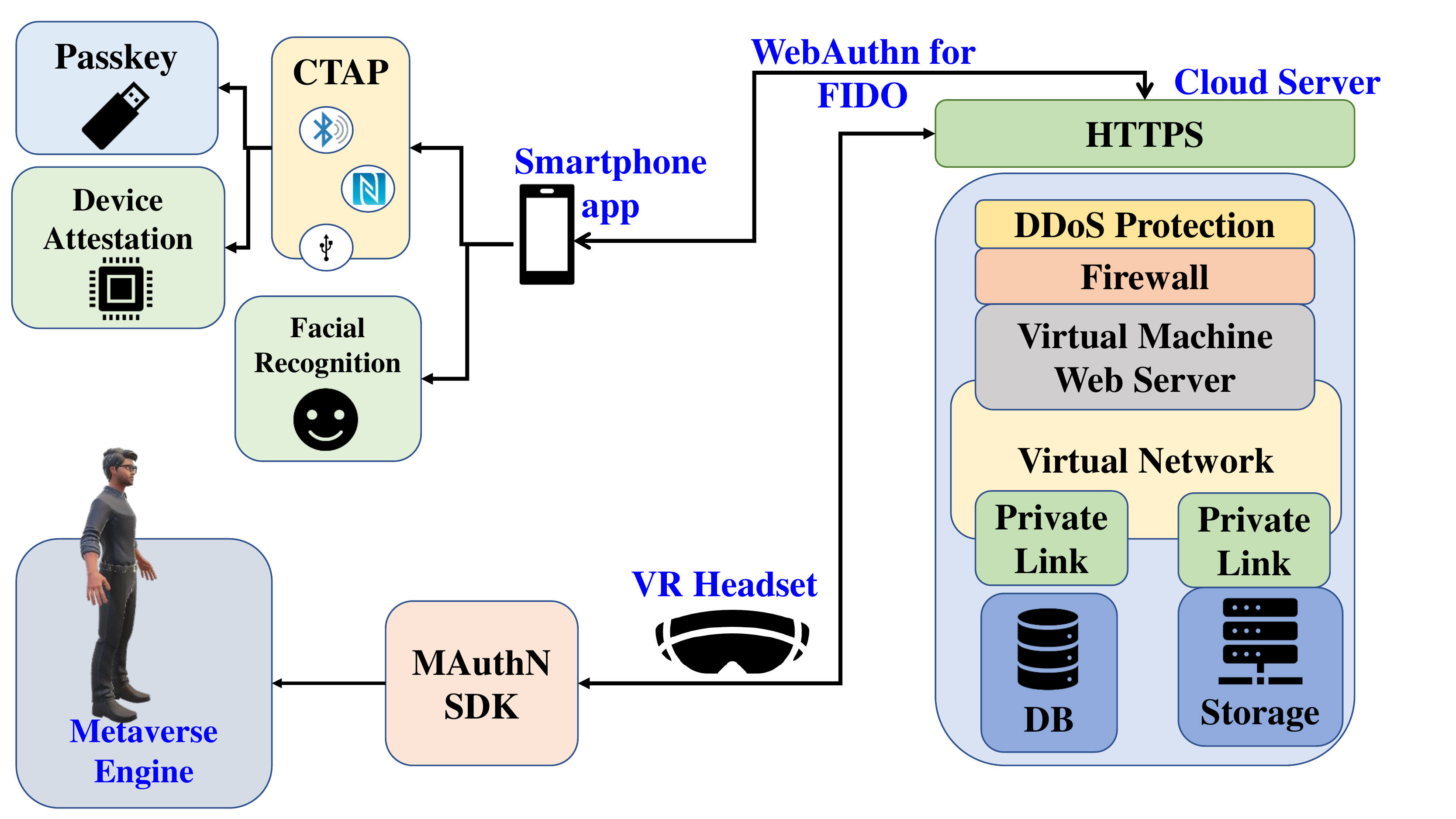}
	\caption{Shows the architecture of the proposed system}\label{fig1:archi}
\end{figure*}

It is evident from the above discussion that there are passwordless multifactor authentication. However, there is no existing Metaverse authentication method implementing physical security keys. Pin Based authentication methods are vulnerable to smudge attacks, thermal imaging attacks and social engineering where one might be made to reveal the PIN. These methods do not make sure that only the authorized users would know the pin, which is not secure. Biometric locks are vulnerable to biometric cloning. Smart locks \cite{ref23} do not have user-defined keys or support biometric authentication. Hence, all these types of security locks have one or the other drawback or vulnerabilities. We believe that Metasecure is the first completely secure multifactor authentication method to implement physical locking system which introduces user-defined keys, seamless key management, access control, triple layer authentication - device attestation, security key and facial recognition to secure digital identity in Metaverse. The workflow of Metasecure is depicted in Figure \ref{fig1:archi}. For authentication using Metasecure, a physical security key is needed. Keys can be managed easily and seamlessly using an online portal by the administrator. Hence, Metasecure also has added security features like remotely wiping the keys in case of a key has been compromised. Further, biometric authentication, a fingerprint compatible physical security key like the YubiKey Bio is to be used. 

\subsection{Design Overview}

The proposed Metasecure implementation is to have a Single Sign On (SSO) service that would request the user to perform authentication and identity assertion operations on a smartphone application. The response for the same would be carried forward to the ‘service provider’ application on which the login is requested. The SSO is to provide suitable Software Development Kit (SDKs) and/or Application Programming Interfaces (APIs). This would be supported on native and console applications, remote devices, IoT devices, Virtual Reality devices, and most Metaverse engines.

The authentication procedure in the proposed system includes the user requesting to authenticate to a specific ‘service provider’. The service provider will forward the request to the server where the proposed system is implemented. The server sends forwards the same request to the smartphone application along with the details of the ‘service provider’. The user can then verify the same on the app and choose to proceed with the authentication process.

\subsection{Authentication}

It is also to be noted that FIDO authentications are extremely fast, yet secure. FIDO uses random cryptographic challenges to mitigate replay attacks. The Relying Party (RP) identity is verified during authentication, as a part of WebAuthn specifications, mitigating all sorts of phishing attacks. It uses the public key cryptosystem to sign a randomly generated challenge which is then securely verified to attest the device or security key.

A security key is a cryptographic authentication device, which is capable of generating RSA keypairs and storing the keys securely such that the private key never leaves the system, making it secure \cite{ref14}. When a key is enrolled against a user, an RSA key pair is generated in the security key. The private key never leaves the security key, while the public key is stored securely on the server, against the user.

\begin{figure*}[!htbp]
	\centering
	\includegraphics[width=0.8\linewidth]{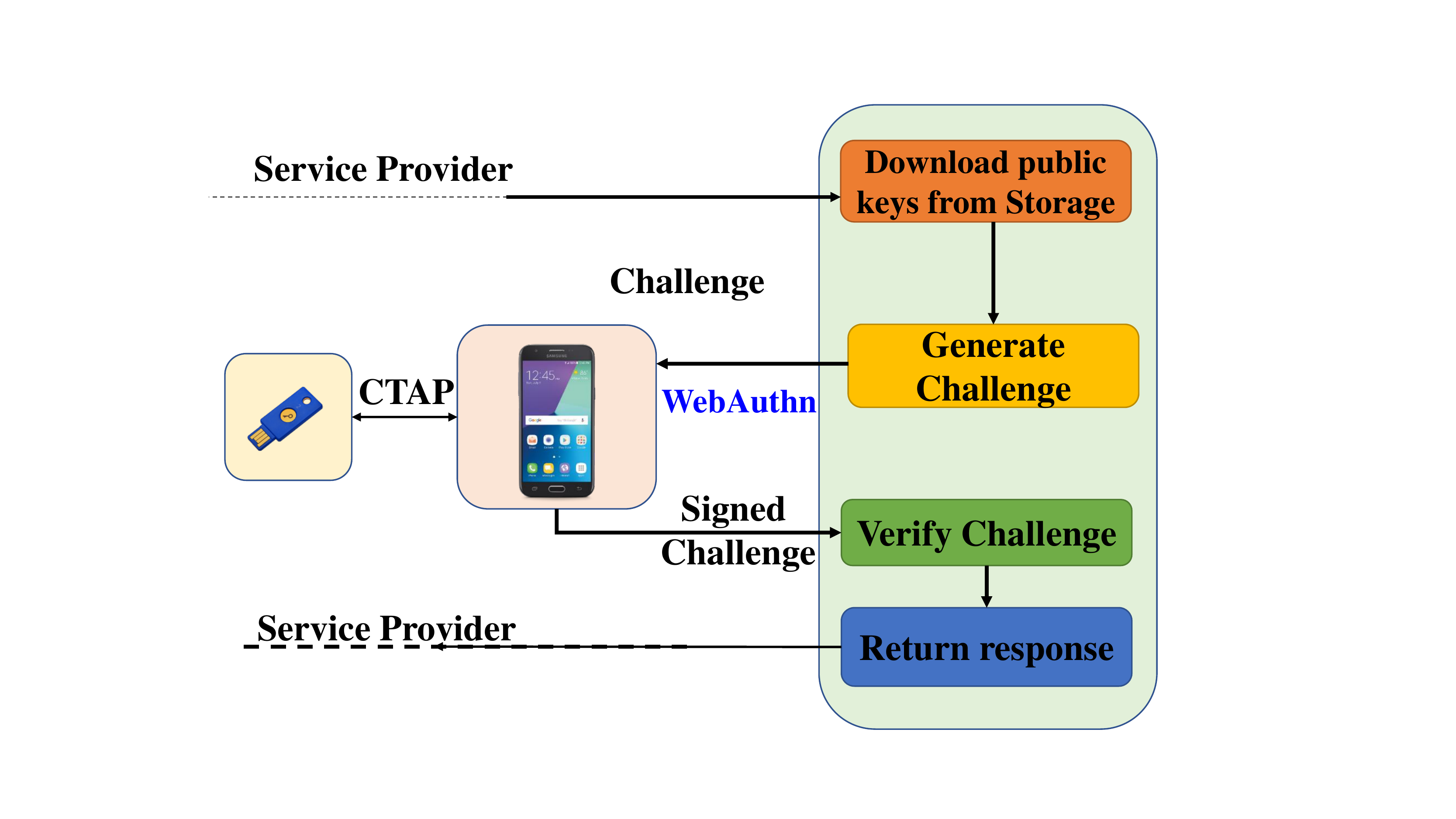}
	\caption{FIDO Authentication workflow}
 \label{fig2:FIDO auth}
\end{figure*}

As the user chooses to authenticate via a security key on the app, the server will generate a random cryptographic challenge for the user, and it will be sent to the user. The user would be requested to connect the security key to the phone via USB, NFC, or BLE. The challenge will then be passed on to the security key where it is signed with the saved private key of the user’s key pair. Then, the same is sent back to the server, where it is verified with the public key to authenticate the user. Figure \ref{fig2:FIDO auth} shows the FIDO authentication workflow.

Metasecure provides the best security when two FIDO compatible devices are enrolled for the user; the recommended ones being the user’s smartphone and a security key.
For asserting the identity of the user, the system proposes the use of facial recognition. Further, the system recommends verifying the input image against presentation attacks. This can be achieved by suitable presentation attack detection algorithms.

\section{Facial Recognition in the Proposed System}
Facial recognition is an important aspect when it comes to biometrics. It involves capturing the image of the user by means of a camera, extracting the facial features, and matching the same against the ones stored in the database against the user’s account. Facial recognition is a branch of computer vision, and it deals with identifying faces in an image or identifying a person from an image. The user’s image is usually matched against a pre-saved image of the user taken during the time of registration. The facial recognition algorithm is to return a confidence value as to how similar the two images are.

\subsection{Automated attack detection}
Presentation attacks provide a substantial amount of risk to biometric systems. A subset of presentation attack determination methods, referred to as liveness detection, involve measurement and analysis of anatomical characteristics or involuntary or voluntary reactions, in order to determine if a biometric sample is being captured from a living subject present at the point of capture. The process of a biometric system detecting a biometric spoof is known as Presentation Attack Detection (PAD). PAD systems utilize a combination of hardware and software technologies to determine whether or not a presented biometric is genuine. A subset of this is liveness detection, which refers to a PAD system’s specific ability to differentiate between human beings and non-living spoofs. 

\begin{figure*}[!htbp]
	\centering
	\includegraphics[width=0.75\linewidth]{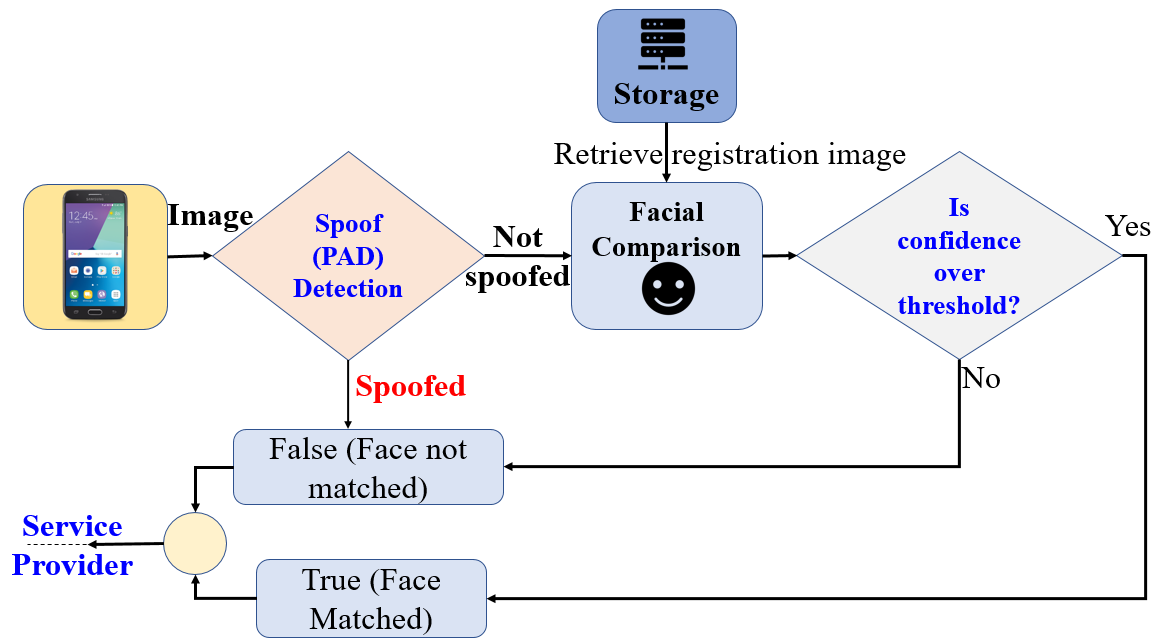}
	\caption{Facial Recognition with PAD}\label{fig3:PAD}
\end{figure*}

Some Presentation Attack Instruments (PAIs)—such as 3D masks, synthetically generated irises, or photographs—can fool a less-secure biometric system. Furthermore, since these attacks occur in the physical world, detection systems often need to include a combination of illumination, sensing, and processing to determine the authenticity of the biometric sample. Therefore, successful PAD often focuses on the interplay between hardware and software to maximize accuracy and usability. PAD is crucial in applications where the combination of security and convenience is a priority. 

\begin{table}[!htbp]
	\caption{Confusion matrix for validation }
	\centering
	\label{tab3:Confusion Matrix}       
	\begin{center}
	\begin{tabular}{|l|c|c|}
		\hline
		\textbf{Accuracy=99.4\%, Loss=0.19} & \textbf{Not spoof} & \textbf{Spoof}\\
		\hline
		Not Spoof &99.2\% & 0.8\% \\
		\hline
		Spoof & 0\% & 100\% \\
		\hline
		F1 Score & 1.00 & 0.99 \\
		\hline
	\end{tabular}
\end{center}
\end{table}

\begin{table}[!htbp]
	\caption{Confusion matrix for testing}
	\centering
	\label{tab4:Confusion Test}       
	\begin{center}
	\begin{tabular}{|l|c|c|c|}
		\hline
		\textbf{Accuracy=98.13\%} & \textbf{Not spoof} & \textbf{Spoof} & \textbf{Uncertain} \\
		\hline
		Not spoof & 98.6\% & 1.4\% & 0\%\\
		\hline
		Spoof & 3.0\% & 97.0\% & 0\%\\
		\hline
		F1 Score & 0.99 & 0.97 & \\
		\hline
	\end{tabular}
\end{center}
\end{table}

This is especially true in automated identification and authentication scenarios — such as physical access control, travel facilitation, payments, or online identity verification— where it may be inefficient, insufficient, or unworkable to have a qualified person manually ensure the authenticity of the presented biometric. time to market, and technology ecosystem support. Figure \ref{fig3:PAD} shows the pipeline of the proposed Facial Recognition system with PAD.

The confusion matrix for validation is given in Table \ref{tab3:Confusion Matrix}. Table \ref{tab4:Confusion Test} shows the confusion matrix for model testing and a transfer learning model has been designed on MobileNet v2 for Presentation Attack Detection. The model achieves a training loss of 0.19 and provided a 99.4\% of validation accuracy and a 98.13\% of testing accuracy.


\section{Metasecure Authentication Mechanism}

Metasecure has been successfully deployed on a fork of VRSpace Metaverse running on Babylon Metaverse engine and can be compatible with other existing or upcoming Metaverse. It is implemented in the following way:
\begin{enumerate}
    \item The user attempts to log in to the Metaverse from any device.
    \item The request is sent to the user’s smartphone.
    \item The user attempts device attestation with the security key (Yubico Key) for Cryptographic authentication.
    \item The user confirms the device he is logging in with.
    \item The user verifies his identity by facial recognition.
    \item The user is now successfully logged in and is able to choose an avatar and enter a world.
\end{enumerate}

\begin{figure*}[!htbp]
	\centering
	\includegraphics[width=0.8\linewidth]{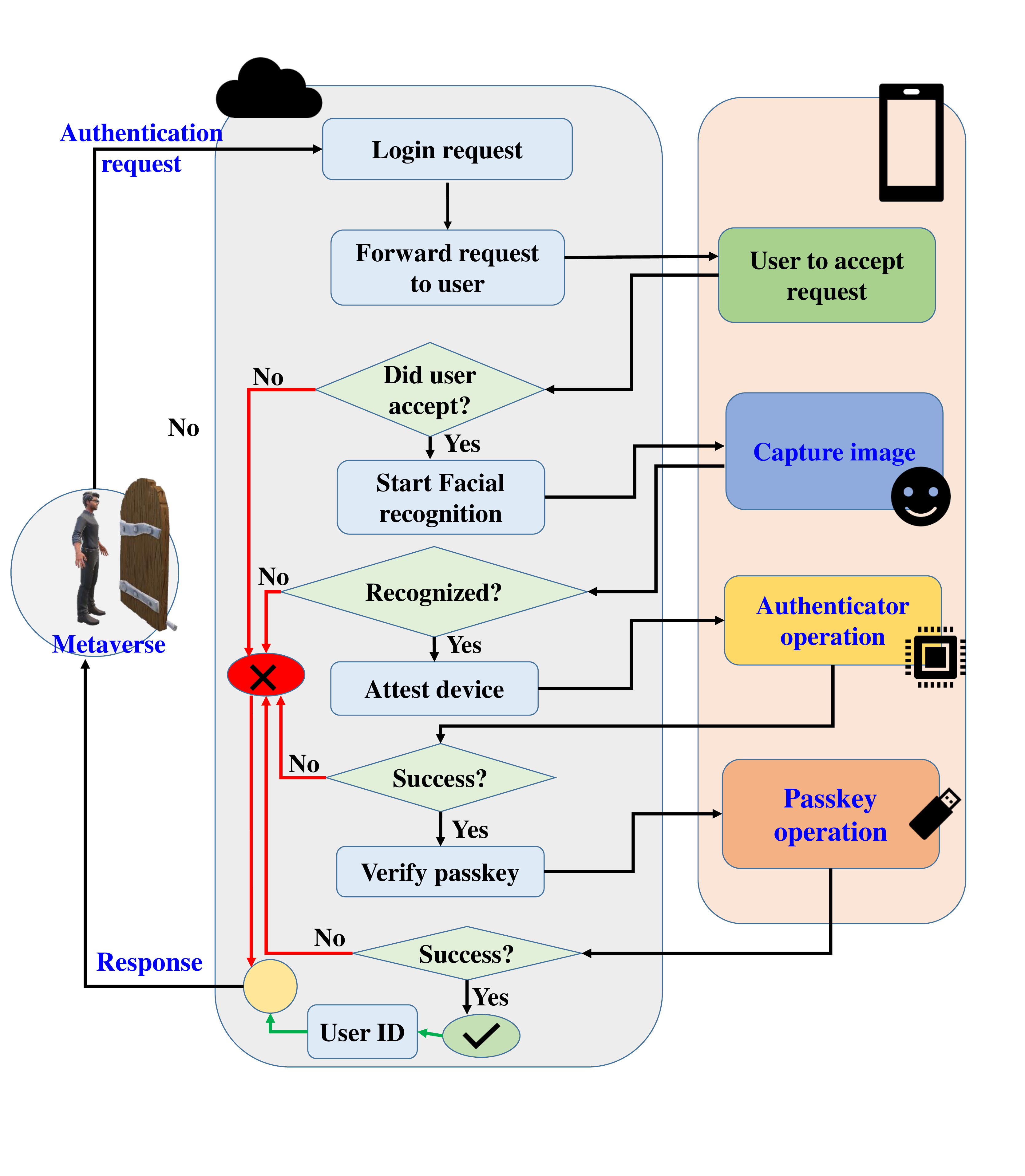}
	\caption{Metasecure Workflow}\label{fig5:Metasecure Work}
\end{figure*}

It is to be noted that the ID of the user, once logged in, would be fixed to his actual identity as registered on the Metasecure database. Figure \ref{fig5:Metasecure Work} shows the workflow of Metasecure.

\begin{figure*} [htbp]
\centering
\includegraphics[height=0.35\textwidth]{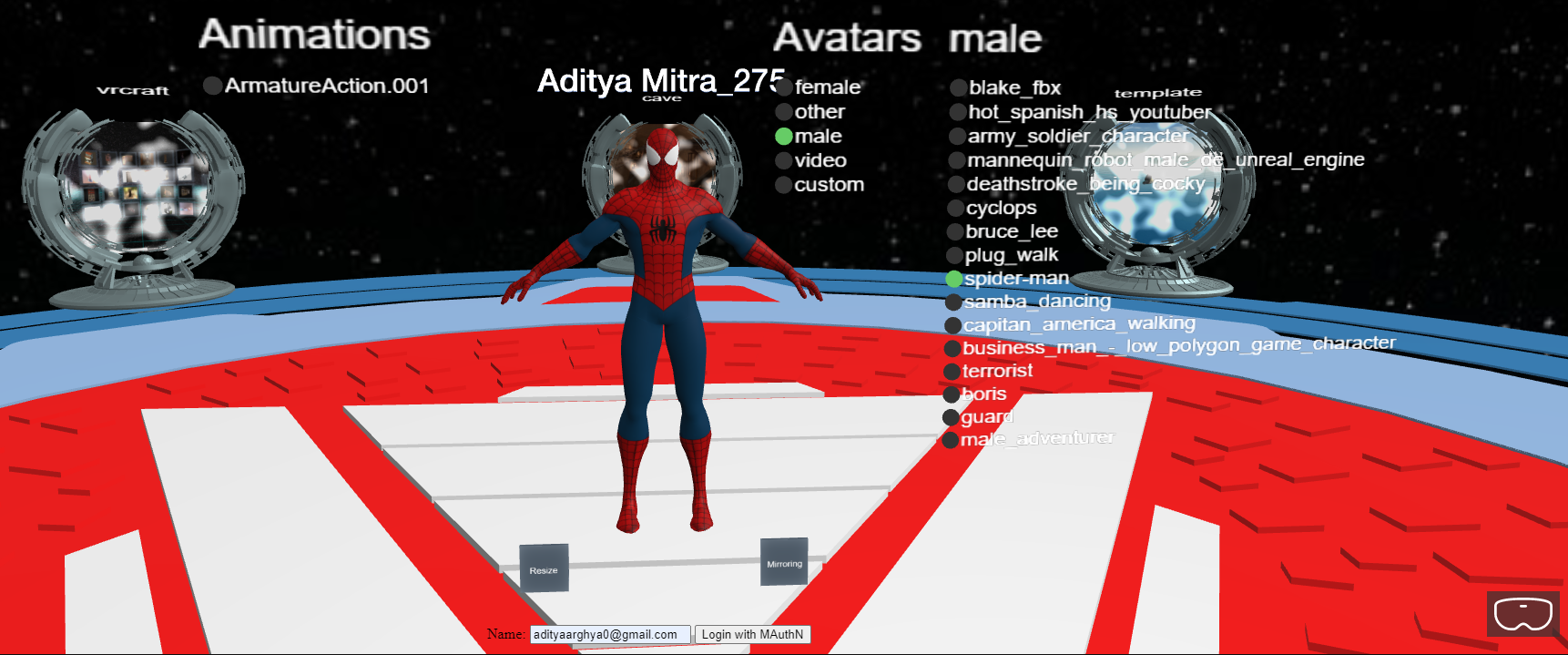}
\caption{Selecting an Avatar in the Metaverse}
\label{FIG:Avatar-Selection}
\end{figure*}

Figure \ref{FIG:Avatar-Selection} shows the proposed login screen. The user is to enter their email ID and perform the authentication operation. Over here the user is allowed to choose any avatar in the Metaverse. 


\begin{figure*} [htbp]
\centering
\includegraphics[height=0.35\textwidth]{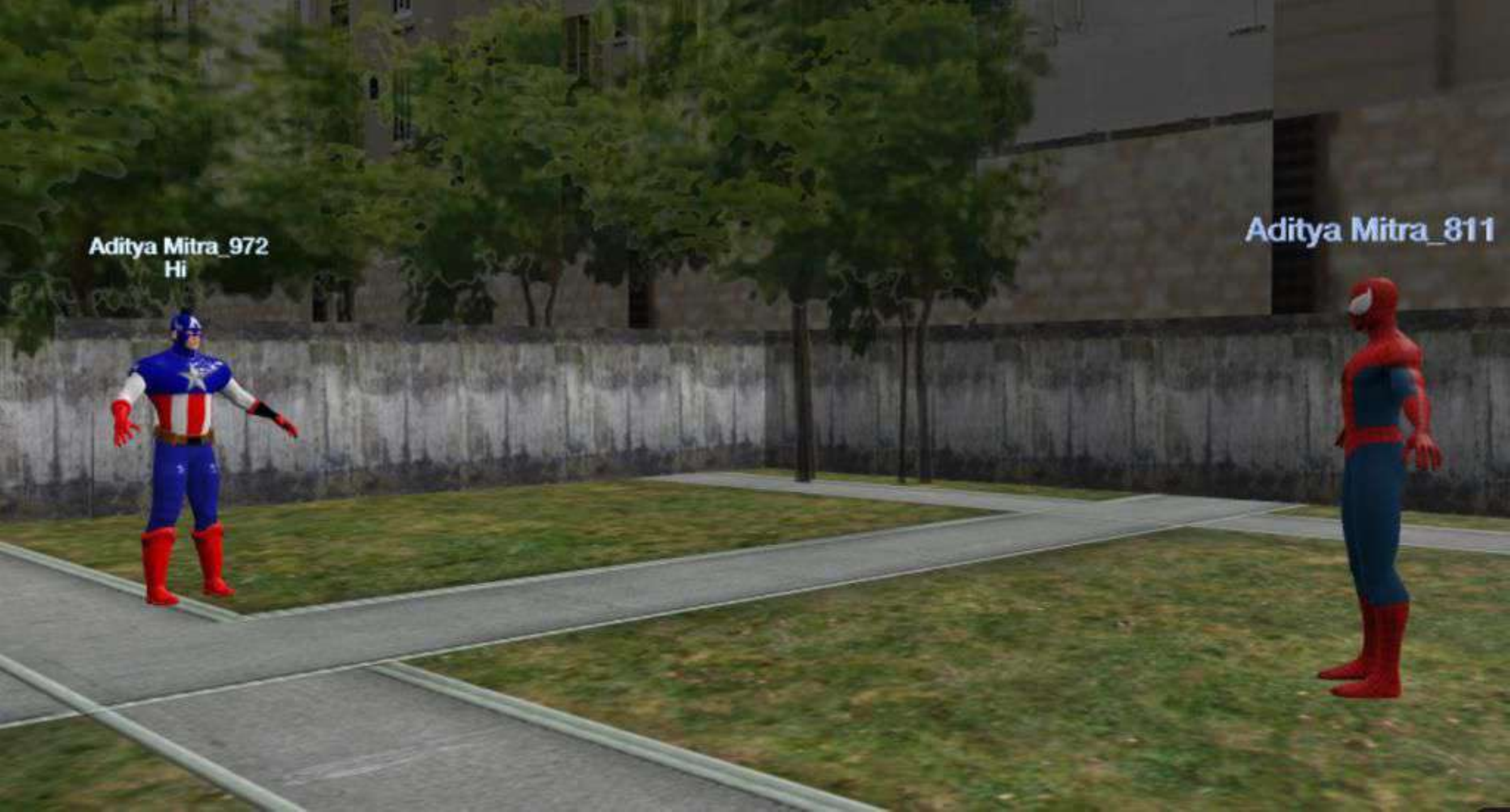}
\caption{Multiple Avatars logged into a world in the Metaverse}
\label{FIG:Metakey-World-Example}
\end{figure*}

\begin{figure}[!htbp]
	\centering
	\includegraphics[width=0.6\linewidth]{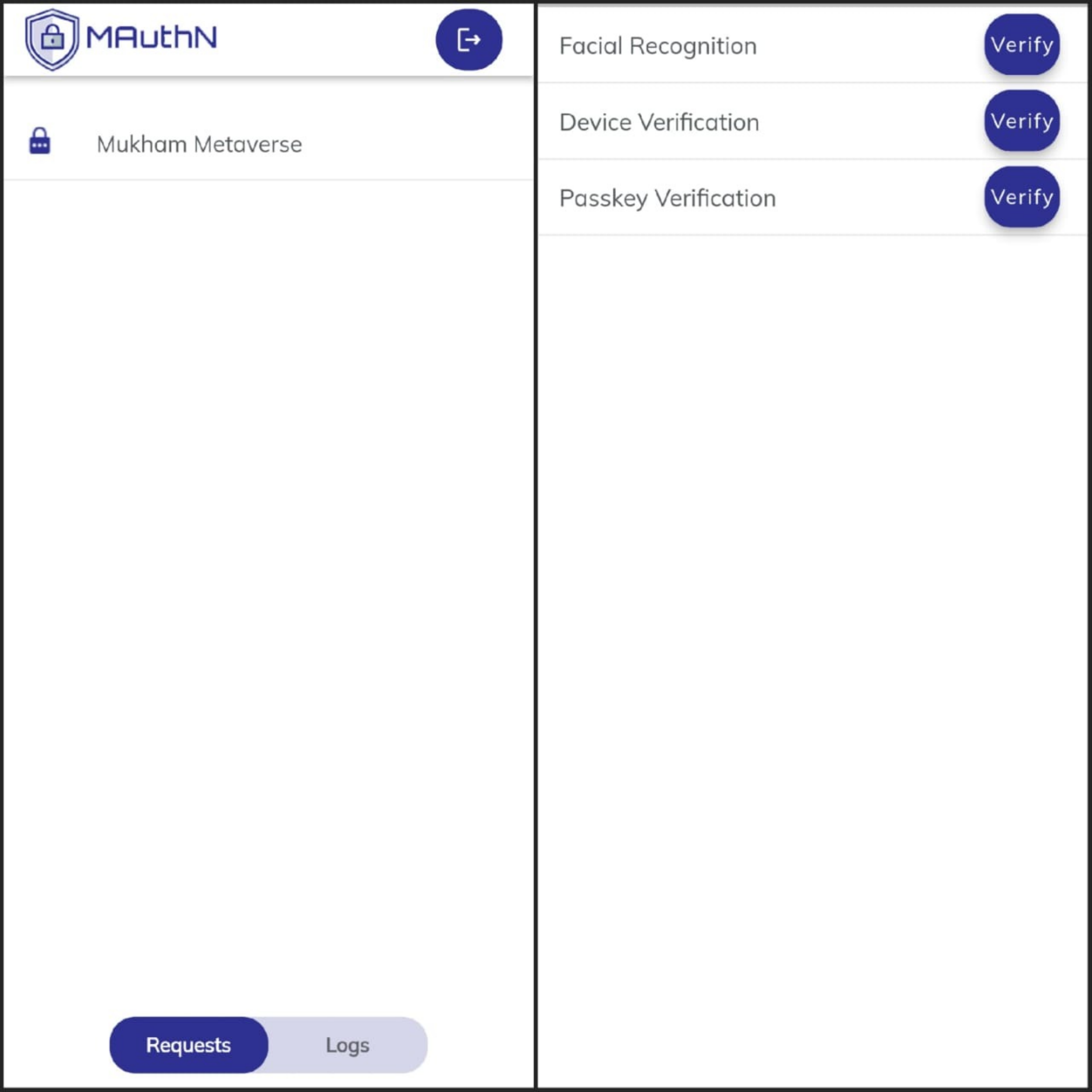}
	\caption{Request on the smartphone app (Left), Multi-layer authentication (Right)}
 \label{fig7:SMARTPHONE APP}
\end{figure}

Figure \ref{fig7:SMARTPHONE APP} shows the request on the smartphone app and the three important authentication steps, namely facial recognition, device verification and security key verification.

Figure \ref{FIG:Metakey-World-Example} shows that once authorized from the smartphone app, the name is fetched directly from the database. The user is not allowed to change their names as it might result in conflicting identity in the Metaverse. Figure \ref{FIG:Metakey-World-Example} also shows multiple avatars entering and navigating the world where they can interact with other users of the Metaverse.



\section{Experimental setup and deployment}
Metasecure authentication servers are deployed on an Azure B2s Virtual machine having 2 vCPUs and 4 GiB RAM. The database is maintained on an Azure SQL server. The user images and the cryptographic data are stored on Azure Blob storage. The backend is written in Python and deployed using Apache2.

The Facial Recognition API is hosted on an AWS EC2 Instance with Apache2. The mobile app is designed and programmed using Flutter. The Metaverse is designed on the Babylon Metaverse engine, as a fork of the open source VRSpace \cite{ref16} project. Further features have been added, which include Indic- languages translation for breaking the language barrier among users. It has been deployed on an AWS EC2 Instance (Virtual Machine) with Spring Boot.

\subsection{Metasecure vs Traditional Login Systems}
Metasecure is a proposed authentication system that overcomes the drawbacks of the current authentication systems. Compared to conventional multi-modal biometrics, Metasecure implements an extensive PAD system to block any attempts for any unauthorized sign ins. Compared to traditional OTPs and TOTPs, Metasecure uses a completely different architecture that focuses more on passwordless authentication.

For normal password based authentication, the password is generally hashed using algorithms like MD5, SHA1, SHA256, SHA512 and stored on the database. It is vulnerable to user side attacks like social engineering, phishing, vishing and so on. If the passwords database on the other hand is breached, it is vulnerable to attacks involving de-hashing the passwords, for example with the help of rainbow tables.

For every password authentication, it needs to be hashed and compared with the one against the database, meaning the time to check a password incorporates the time needed to hash it. Considering a 49 character password versus passwordless authentication, the time taken for each can be demonstrated over 5 test cases as in Table \ref{tab4:Confusion Test} \cite{ref15}.

\begin{table*}[!htbp]
	\caption{Time comparison of various facial recognition models}
	\centering
	\label{tab5:comparison of various facial recognition models}       
	\begin{center}
	\begin{tabular}{|l|c|c|c|c|c|c|c|c|}
		\hline
		\textbf{Test Cases} & \textbf{VGG-Face} & \textbf{Facenet} & \textbf{Facenet512} & \textbf{OpenFace} & \textbf{DeepFace} & \textbf{DeepID} & \textbf{ArcFace} & \textbf{SFace}\\
		\hline
		Test case 1 (CPU) & 1140 ms & 200 ms & 282 ms & 176 ms & 301 ms & 145 ms & 489 ms & 131 ms \\
		\hline
		Test case 2 (GPU) & 249 ms & 157 ms & 172 ms & 135 ms & 234 ms & 126 ms & 143 ms & 130 ms\\ 
		\hline
		Test case 3 (CPU) & 1150 ms & 282 ms & 282 ms & 177 ms & 298 ms & 145 ms & 500 ms & 133 ms \\
		\hline
		Test case 4 (GPU) & 257 ms & 193 ms & 183 ms & 145 ms & 238 ms & 130 ms & 166 ms & 132 ms \\
		\hline
		Average (CPU) & 1145 ms & 241 ms & 282 ms & 176.5 ms & 299.5 ms & 145 ms & 494.5 ms & 132 ms\\
		\hline
		Average (GPU) & 256 ms & 175 ms & 177.5 ms & 140 ms & 236 ms & 128 ms & 154.5 ms & 131 ms\\
		\hline
	\end{tabular}
\end{center}
\end{table*}

Comparing multiple authentication models, including biometrics, multi-modal biometrics, regular passwords and so on, it can be concluded that Metasecure is a faster, secure, novel and seamless than current existing  login systems for Metaverse. 

\section{Conclusion and Future Work}

This paper presents a standard that allows FIDO2 compliant physical security keys to be used with device attestation and facial recognition to make Metaverse secure. It discusses the merits of leveraging Metasecure for Metaverse authentication and identity management. The proposed system would be instrumental in mitigating identity thefts, abusing the authentication systems. This would also be instrumental in mitigating cyberstalking, harassment and related issues on the Metaverse. Metasecure can  be deployed on any Metaverse engine using the wide range of available SDKs seamlessly and can protect the user’s identity and assets in the digital world. The paper also discusses advantages of using Metasecure over other passwordless and password based authentication methods. It has been made sure that not only authentication is secure but also user identity is thoroughly verified to implement secure interactions. According to the best of our knowledge, Metasecure is the first authentication module  can be implemented to implement FIDO2 based security, enabling seamless access control where user can be authorized and unauthorized from accessing digital assets in Metaverse remotely.

As future work, this work can be further improved for internal Metaverse features such as to remove eavesdropping, remove financial frauds using NFTs and many more may be proven beneficial.

\section*{Acknowledgments}
We thank Dr. S.V. Kota Reddy, Vice Chancellor and Dr. Hari Seetha, Director, COE, Artificial Intelligence and Robotics (AIR), VIT-AP University for motivating and helping us to build this project. A special thanks to the team members of AIR, VIT-AP University.


\end{document}